\documentclass[conference,10pt]{IEEEtran}
\IEEEoverridecommandlockouts
\usepackage{listings}
\usepackage{amsmath,amssymb,amsfonts}
\usepackage{algorithmic}
\usepackage{graphicx}
\usepackage{textcomp}
\usepackage[usenames,dvipsnames]{xcolor}
\usepackage{tcolorbox}
\def\BibTeX{{\rm B\kern-.05em{\sc i\kern-.025em b}\kern-.08em
    T\kern-.1667em\lower.7ex\hbox{E}\kern-.125emX}}

\usepackage{hyperref}

\usepackage{capt-of}
\usepackage{listings}
\usepackage{subcaption}
\usepackage[style=ieee,maxnames=1,minnames=1,backend=biber,bibstyle=ieee,citestyle=numeric-comp]{biblatex}
\usepackage{url}
\usepackage{tcolorbox}
\usepackage{rotating}
\usepackage{textpos}
\newcommand\PlaceText[4]{%
    \begin{textblock*}{1.1#1}(2pt,#2)
        \begin{turn}{-90}
            \begin{tcolorbox}[width=#3,height=0.90#1, standard jigsaw, opacityback=0,boxsep=-2pt]
            \centering
            \scriptsize
            #4
            \end{tcolorbox}
        \end{turn}
    \end{textblock*}}
\begin{document}
\renewcommand{\thelstlisting}{\arabic{lstlisting}}

\definecolor{codegreen}{RGB}{53,114,0}
\definecolor{codegray}{rgb}{0.5,0.5,0.5}
\definecolor{backcolour}{RGB}{255,250,234}
\definecolor{codered}{RGB}{157,0,6}
\definecolor{codeblue}{RGB}{22,102,176}
\definecolor{codeaqua}{RGB}{0,112,114}
\definecolor{codepurple}{RGB}{164,56,192}
\definecolor{codeHighlight}{RGB}{227,242,253}
\lstdefinelanguage{C_more}{
    keywords=[1]{cudaGraphNode_t, int, void, cudaKernelNodeParams, for, unsigned},
    keywordstyle=[1]\color{blue},
    keywords=[2]{NULL},
    keywordstyle=[2]\color{purple},
    keywords=[3]{cudaGraphCreate, cudaGraphAddKernelNode, cudaGraphInstantiate, cudaStreamCreateWithFlags, cudaGraphUpload, kernel, cudaGraphLaunch},
    keywordstyle=[3]\color{NavyBlue},
    comment = [l]{//},
    commentstyle=\color{OliveGreen},
    tabsize=2,
    frame=tb,
    numbers=left,
    numbersep=1em,
    numberstyle=\tiny\color{codegray},
    xleftmargin=2em,
    backgroundcolor=\color{backcolour},
    basicstyle=\scriptsize,
    breakatwhitespace=false,
    breaklines=true,
    captionpos=t,
    keepspaces=true,
    showspaces=false,
    showstringspaces=false,
    showtabs=false,
    framexleftmargin=0.1em,
}
\lstdefinestyle{CEE}{
    language=C_more,
}
\lstdefinestyle{CEE_base}{
    language=C_more,
}
\lstdefinestyle{CEE_highlight}{
    language=C_more,
    backgroundcolor=\color{codeHighlight},
    frame=none,
    framesep=0pt,
    framexleftmargin=1.5em,
}

\newcounter{obsno}
\definecolor{textHighlight}{RGB}{227,242,253}
\newcommand{\observation}[1]{
    \begin{tcolorbox}[width=\linewidth, colback=textHighlight,left=2pt,right=2pt,top=2pt,bottom=2pt]
        \textbf{\MakeUppercase{Observation} \refstepcounter{obsno}\Roman{obsno}:} #1
    \end{tcolorbox}
}

\title{
Boosting Performance of Iterative Applications on GPUs: Kernel Batching with CUDA Graphs
\thanks{This work is supported by the EuroHPC Joint Undertaking project Plasma-PEPSC, project no. 101093261.}
}
\author{\IEEEauthorblockN{Jonah Ekelund, Stefano Markidis, Ivy Peng}
\IEEEauthorblockA{\textit{KTH Royal Institute of Technology, Sweden}}
\{jonakek, markidis, bopeng\}@kth.se
}

\maketitle

\begin{abstract}
Graphics Processing Units (GPUs) have become the standard in accelerating scientific applications on heterogeneous systems. However, as GPUs are getting faster, one potential performance bottleneck with GPU-accelerated applications is the overhead from launching several fine-grained kernels. \texttt{CUDA~Graph} addresses these performance challenges by enabling a graph-based execution model that captures operations as nodes and dependence as edges in a static graph. Thereby consolidating several kernel launches into one graph launch. We propose a performance optimization strategy for iteratively launched kernels. By grouping kernel launches into iteration batches and then unrolling these batches into a \texttt{CUDA~Graph}, iterative applications can benefit from \texttt{CUDA~Graph} for performance boosting. We analyze the performance gain and overhead from this approach by designing a skeleton application. The skeleton application also serves as a generalized example of converting an iterative solver to \texttt{CUDA~Graph}, and for deriving a performance model. Using the skeleton application, we show that when unrolling iteration batches for a given platform, there is an optimal size of the iteration batch, which is independent of workload, balancing the extra overhead from graph creation with the performance gain of the graph execution. Depending on workload, we show that the optimal iteration batch size gives more than $1.4\times$ speed-up in the skeleton application. Furthermore, we show that similar speed-up can be gained in Hotspot and Hotspot3D from the Rodinia benchmark suite and a Finite-Difference Time-Domain (FDTD) Maxwell solver.
\end{abstract}
\begin{IEEEkeywords}
CUDA~Graph, GPU Task Execution Model, GPU Performance Optimization, Hotspot3D, FDTD
\end{IEEEkeywords}

\section{Introduction}
Graphics Processing Units (GPUs) have emerged as a major workforce for a wide range of workloads, ranging from deep learning~\cite{bengio2017deep} to scientific applications, such as molecular dynamics simulations~\cite{andersson2022breaking}, computational fluid dynamics~\cite{karp2023large} and quantum computer simulators~\cite{faj2023quantum}. However, as the GPU's computational capabilities in terms of FLOPS/s increase, while the execution time of the computational kernels decreases, the cost of kernel launching remains approximately constant. This is due to different factors not directly dependent on improved GPU micro-architectures. These factors include interaction with the OS, driver and API constraints (which have not evolved as fast as the hardware), CPU-GPU communication overheads, and fixed latency in launch setup. To address this fundamental limitation due to the kernel launching cost, Nvidia introduced \texttt{CUDA~Graph} in CUDA 10, released in 2018. \texttt{CUDA~Graph} consolidates multiple kernel launches into a single graph launch, reducing the overall launch overhead~\cite{pramodramaraoCUDA10Features2018}. This graph-based approach helps minimize the repetitive setup cost by batching and organizing kernels. It reduces the dependence on CPU-driven launches, fundamentally limiting the application performance. Instead, it allows multiple kernels to be launched as a unified workload. In this work, we develop a performance model, port HPC applications to use \texttt{CUDA~Graph}, and evaluate its performance benefits.

 A large number of scientific HPC applications are iterative in nature: they include a loop where the solver kernel is offloaded to the GPU at each iteration. Examples of such applications are all the applications based on time-stepping, such as electromagnetic solvers, particle pushers, Computational Fluid Dynamics (CFD) solvers, to cite a few notable examples. Iterative applications are ideal to use \texttt{CUDA~Graph} capabilities: the kernel launches at successive iterations can be grouped in \textit{batches} and use a single launch. In this work, we develop a performance model that shows that there should be an optimal size of the \textit{batches} which balances performance improvements with the added overhead of graph creation. We then design and develop \texttt{CUDA~Graph} ports of a skeleton version of an iterative application, which then is used to validate a performance model, the HotSpot application, which models the thermal diffusion on a CPU, and an electromagnetics application, based on the Finite Difference Time Domain (FDTD) numerical scheme. Using the skeleton application, we perform a performance characterization of \texttt{CUDA~Graph} performance on Nvidia A100 GPU, varying the batch size, e.g. the number of grouped kernels, the total number of iterations, and the total number of CUDA threads. The results are then validated on a  Grace-Hopper system.    

\begin{figure*}[bt]
    \centering
    \begin{minipage}{0.8\linewidth}
    \centering
    \includegraphics[width=\linewidth]{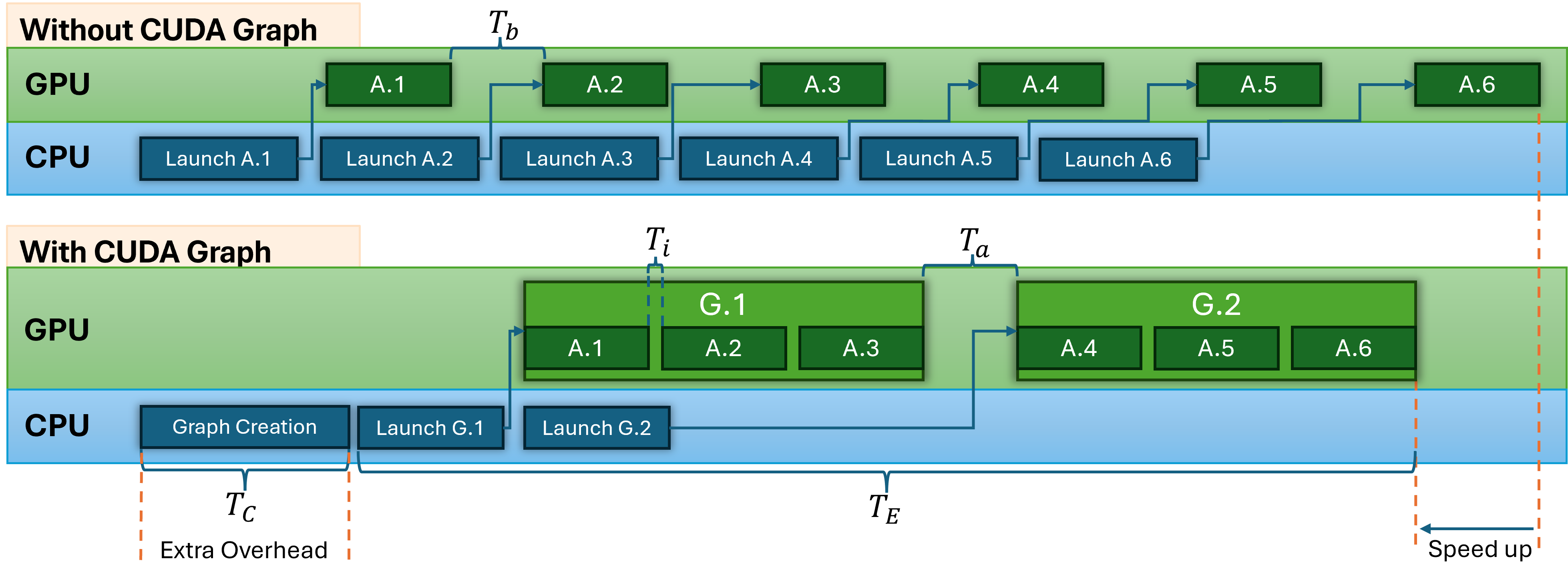}
    \caption{A schematic representation of the iteration batch unrolling strategy presented in this paper. }
    \label{fig:cg-launch-cost}
    \vspace{-1em}
    \end{minipage}
\end{figure*}

The goal of this work is to introduce a methodology to adopt \texttt{CUDA~Graph} in iterative applications, with an associated performance model, and evaluate its performance improvement. Overall, we find that by applying this methodology, we can achieve performance improvements of up to 40\%. The main contributions of this work are the following:
\begin{itemize}
    \item The development of a methodology to use \texttt{CUDA~Graph} in iterative applications, by grouping the kernel launches into batches, and using \texttt{CUDA~Graph}, see Fig.~\ref{fig:cg-launch-cost}.
    \item The design of a skeleton version of an iterative application using \texttt{CUDA~Graph} and a performance model to understand the performance benefits of the proposed approach.
    \item We provide a characterization study and a performance model of \texttt{CUDA~Graph} in iterative applications, varying key parameters like the batch size and number of iterations. The performance model and benchmarking can be used to estimate the parameters for the largest speed-up.
    \item We detail all the steps to port an iterative application to use \texttt{CUDA~Graph} with code snippets and open-source code.
\end{itemize}

\section{Background}
\label{sec:bg}
Today, GPUs are a common component in both personal computers and supercomputers. Originally designed to process computer graphics, GPUs are specialized computing units built for high data throughput by processing multiple data points with the same instruction in parallel, i.e. Same Instruction Multiple Data (SIMD). This is useful in scientific applications where the same calculation is performed on multiple data inputs, such as simulations and neural networks.

CUDA is a programming model and a platform for using Nvidia GPUs as general computing devices in applications~\cite{CUDAZoneLibrary}. In CUDA, work for the GPU is defined as kernels launched to the GPU. The kernel is executed in several threads specified at launch time, where each thread executes the same instruction but can operate on different data. CUDA also has APIs for other operations, such as allocating memory in the GPU for use in the application and copying data to and from the GPU.

In a task-graph execution model, operations are defined as \textbf{nodes} and dependencies between operations are explicitly captured as \textbf{edges} in a graph. \texttt{\texttt{CUDA~Graph}} was introduced in CUDA 10 to enable such task graph execution model, where nodes represent operations including kernel launches, CPU function calls, \texttt{memcpy} or \texttt{memset}, allocation/de-allocation. When a sequence of kernels is captured in one static graph, the CPU side only needs to launch the graph once, instead of paying the launch overhead for each kernel as in native CUDA programs. Fig.~\ref{fig:cg-launch-cost} contrasts the execution timeline of an iterative application executing with and without \texttt{CUDA~Graph}. The task graph execution model can reduce launch overhead and free up CPU resources in applications, such as Deep Neural Networks or simulations, where several short kernels are relaunched many times~\cite{pramodramaraoCUDA10Features2018}. However, when converting a native CUDA program into the task graph model, extra overhead is required for creating, initializing, and launching graphs. Thus, this overhead requires quantitative consideration when composing the transformation strategy and the benefits of \texttt{CUDA~Graph} in HPC application patterns.

There are two ways of constructing \texttt{CUDA~Graph}. The first technique is called \underline{Stream Capture}. It allows the capture of an existing CUDA stream implementation in a graph. This is useful if the actual operations to be included in the graph are in libraries, such as cuBLAS or cuSPARSE and not directly accessible to the programmer~\cite{nvidiaCUDAGraphs,nvidiaGraphManagement}. Stream Capture can be simple to implement, however, it requires that the streams are synced by inserting CUDA events. A graph created with the stream capture API can also be difficult to update once it has been created~\cite{yuOpenMPCUDAGraphs2020,linEfficientGPUComputation2021,saysConstructingCUDAGraphs2022}. The second approach is \underline{Manual Graph Creation} -- the creation of the graph and adding of nodes to the graph is performed through manual CUDA calls. This gives the developer precise control over the graph creation and all dependencies. This requires the developer to know all the configurations for the operations. It also takes more effort to implement and can be harder to maintain~\cite{yuOpenMPCUDAGraphs2020}. In this work, we have chosen to use manual graph creation due to the added control over the graph this enables.

\section{Methodology}
\label{sec:methodology}
In this section, we first describe our iteration batch unrolling strategy. We then describe the characterization and optimization tests performed on the skeleton application, followed by measurement and error calculations.

\subsection{Iteration Batch Unrolling into a Static Graph}\label{sec:opt-batching}
In many scientific applications, the main computation phase consists of many iterations of a small set of kernels, where a common implementation style is to call the CUDA kernel in a loop for a specified number of iterations, see Listing~\ref{lst:mb-wo-graph}.
\begin{lstlisting}[style=CEE,caption={Baseline implementation of an iterative application.},label=lst:mb-wo-graph]
for (int i = 0; i < kernel_iterations; ++i) {
    kernel<<<grid, block>>>(...);
}
\end{lstlisting}
Constructing a \texttt{CUDA~Graph} with only one or two nodes is unlikely to yield any performance benefit when the kernels need to be executed in sequence. On the other end, creating a \texttt{CUDA~Graph} with a node for each call to launch the kernel would result in an extensively large graph, with a large creation overhead. Therefore, to balance the added overhead of creating the graph, we propose to first group a number of the kernel launches together into an iteration batch. Listing~\ref{lst:mb-wo-graph-batch} shows how this would look in the example in Listing~\ref{lst:mb-wo-graph}. We then unroll this iteration batch into a \texttt{CUDA~Graph} which can be iterated on. This iteration batch unrolling strategy is similar to loop unrolling in compiler optimizations.

\begin{lstlisting}[style=CEE,caption={Example of iteration batching in a traditional implementaiton.},label=lst:mb-wo-graph-batch]
for (int i = 0; i < nr_iter_batches; ++i) {
    for (int j = 0; j < iteration_batch_size; ++j){
        kernel<<<grid, block>>>(...);
    }
}
\end{lstlisting}

By unrolling the iteration batch into a \texttt{CUDA~Graph}, we can reduce the launch overhead by using fewer graph launches. Furthermore, presenting a larger workflow to the runtime can enable the runtime to perform more optimization on the execution. Therefore, we should be able to amortize the cost of graph creation and initialization and gain performance by reducing the overall execution time. The challenge is determining a proper size for the iteration batch while the memory usage and performance are balanced. Based on our measurements, the memory overhead of a \texttt{CUDA~Graph} can be substantial if too many nodes are added.

An example of how the unrolling is implemented using the Manual Graph Creation API to create the graph can be seen in Listing~\ref{lst:mb-graph}. Note that this increases the Lines of Code (LoC) significantly compared to the implementation in Listing~\ref{lst:mb-wo-graph}.

\noindent\begin{minipage}{\linewidth}
\begin{lstlisting}[style=CEE_base,frame=t,caption={Graph implementation of an iterative application.},label=lst:mb-graph]
cudaGraphCreate(graph, NULL);

void *ka_kernel[] = {...}; //Parameters for the kernel
cudaKernelNodeParams np_kernel = {0};
np_kernel.func = (void *)kernel;
np_kernel.gridDim = grid;
np_kernel.blockDim = block;
np_kernel.kernelParams = ka_kernel;
cudaGraphNode_t *last_node = NULL;
int num_dependencies = 0;
\end{lstlisting}
\vspace{-0.95\baselineskip}
\begin{lstlisting}[style=CEE_highlight,firstnumber=11]
for (int i = 0; i < iteration_batch_size; ++i) {
    cudaGraphAddKernelNode(&nodes[i], graph,
        last_node, num_dependencies,
        &np_kernel);
    last_node = &nodes[i];
    num_dependencies = 1;
}
\end{lstlisting}
\vspace{-1.07\baselineskip}
\begin{lstlisting}[style=CEE_base,firstnumber=18,frame=none,escapechar=\%]

cudaGraphInstantiate(&exec_graph, graph,
    cudaGraphInstantiateFlagDeviceLaunch);

cudaStreamCreateWithFlags(&stream_for_graph,
    cudaStreamNonBlocking);
cudaGraphUpload(exec_graph, stream_for_graph);
%
\end{lstlisting}
\vspace{-0.95\baselineskip}
\begin{lstlisting}[style=CEE_highlight,frame=b,firstnumber=26]
// kernel_iterations have to be divisable
// by iteration_batch_size
int nr_iter_batches = kernel_iterations/iteration_batch_size;
for (int i = 0; i < nr_iter_batches; ++i) {
    cudaGraphLaunch(exec_graph, stream_for_graph);
}
\end{lstlisting}
\PlaceText{\linewidth}{-18.35em}{5.8em}{Building}
\PlaceText{\linewidth}{-6.4em}{5.9em}{Launching}
\end{minipage}
\subsection{Performance Modeling}
\label{sec:r-mb}
As shown in Fig.~\ref{fig:cg-launch-cost}, harnessing \texttt{CUDA~Graph} requires a trade-off between the performance gain from the graph execution and the increased overhead from graph creation. Therefore, when evaluating the execution performance of unrolling the iteration batch into a \texttt{CUDA~Graph} we need to consider both the graph creation overhead $T_C$ and the execution time $T_E$
\begin{equation}\label{eq:T_total}
    T = T_C + T_E.
\end{equation}
Based on how nodes are added in a linear sequence in Listing~\ref{lst:mb-graph}, see the highlighted Building section, the graph creation should be described by a linear equation
\begin{equation}\label{eq:creation_overhead}
    T_C =  k_c S + b_c,
\end{equation}
where $S$ is the iteration batch size, $k_c$ and $b_c$ are the linear coefficients. From Fig.~\ref{fig:cg-launch-cost} we can see that the execution $T_E$, can be described as the sum of the time to launch the first graph $T_l$, the execution time of one graph $T_{eg}$ multiplied by the number of iteration batches to launch $I$ and the total time between the graph executions $T_a(I-1)$:

\begin{equation} \label{eq:exec_time}
  T_E = T_l +
    \underbrace{(T_kS + T_i(S-1))}_{T_{eg}}
    I +
    T_a(I-1),
\end{equation}
here $T_{eg}$ is expanded as a sum between the execution time of all the kernels [$T_kS$] and the time between kernel executions in the graph [$T_i(S-1)$]. For a given number of total kernel executions ($I_k$) the relationship between the number of iteration batches ($I$) and iteration batch size ($S$) is described by
\begin{equation} \label{eq:graph_executions}
    I = \frac{I_k}{S}.
\end{equation}
Using this in Eq.~\ref{eq:exec_time} and rearranging we obtain
\begin{equation}\label{eq:exec_time_2}
    T_E = \frac{I_k(T_a-T_i)}{S} + I_k(T_k + T_i) - T_a + T_l.
\end{equation}

Early exploration indicated that the time between graphs is longer than the time between kernels in the graph ($T_a > T_i$) and that $T_a$ does not change with the graph size. Furthermore, as the graph is uploaded to the GPU before it is launched, line 24 in Listing~\ref{lst:mb-graph}, $T_l$ should also be independent of the graph size. According to to Eq.~\ref{eq:exec_time_2} we would therefore expect to obtain the best execution performance when the iteration batch size is equal to the total number of kernel executions ($S=I_k$). If we then assume that switching from one workload (graph or kernel) launched to the GPU is approximately the same ($T_a \approx T_b$ in Fig.~\ref{fig:cg-launch-cost}), executing the kernels as part of a graph will give a speedup.

However, this has to be balanced against the additional graph creation overhead from Eq.~\ref{eq:creation_overhead}. This suggests that there should be an optimal size for the iteration batch independent of the kernel execution ($T_k$). If we define $a = I_k(T_a-T_i)$ and $b = I_k(T_k + T_i) - T_a + T_l$, we can write Eq.~\ref{eq:exec_time_2} as
\begin{equation} \label{eq:exec_time_fit}
    T_e = \frac{a}{S} + b
\end{equation}
where the coefficient $a$ relates to the speed-up obtained by increasing the number of nodes in the graph. For large $I_k$ we have that  $I_k(T_k + T_i) >> -T_a + T_l$. $T_a$ and $T_l$ can thereby be ignored and $b$ will then represent a lower bound on the possible execution time where all kernels are part of one large graph. 

\observation{\textit{The iteration batch unrolling strategy can achieve performance gains when the time to switch to the next kernel inside the graph ($T_i$) is lower than the time to switch from one launched kernel to the next launched kernel ($T_a$).}}

We identify the value of the iteration batch $S$, such that Eq.~\ref{eq:T_total} is minimized, to optimize the performance.  However, as the total number of kernel executions ($I_k$), the number of iteration batches ($I$) and the iteration batch size ($S$) all are limited integer numbers, Eq.~\ref{eq:graph_executions} also limits the possible iteration batch sizes for a given total number of kernel executions.

\subsection{Experimental Setup} \label{sec:setup}

To analyze the graph creation overhead $T_C$ and graph execution $T_E$, we designed a \textbf{skeleton version of an iterative application}\footnote{Skeleton application: \url{https://github.com/Jonah-E/IterApp}}, providing a generalized example of how an iterative application can be converted to use \texttt{CUDA~Graph}. The computational workload in the skeleton application is a simple vector multiplication kernel, multiplying a constant value to a vector. Through a command line interface, the execution of the application can be changed to run either with or without \texttt{CUDA Graph}. This enables a comparison between the baseline GPU implementation and the graph implementation. Furthermore, the size of the iteration batch, the number of iteration batches, and the number of threads per kernel can be changed through the command line interface. When running the baseline GPU implementation, the graph launch is replaced with a for-loop launching all the kernels in an iteration batch, similar to what is shown in Listing.~\ref{lst:mb-wo-graph-batch}. Using the skeleton application, we test the overhead and performance of a \texttt{CUDA~Graph} when increasing the size of the iteration batch (for the graph version, this is the same as the number of nodes in the graph) while keeping the total number of kernel executions constant. By measuring the graph creation time and the execution time, we evaluate Eqs.~\ref{eq:creation_overhead} and~\ref{eq:exec_time_fit}.

To assess how the memory footprint of the created graph scales with the iteration batch size, we measure the memory usage of the skeleton application for the increasing size of the graph. This is done by calling CUDA Systems management Interface (\texttt{nvidia-smi}) from within the application after the execution of the graph, but before the graph is destroyed.

Finally, to find the optimal graph size we use the skeleton application and measure the graph creation and total execution time while increasing iteration batch size and keeping the number of required kernel executions ($I_k$) constant. Using the optimal graph size, we can then evaluate the speed-up of the skeleton application, Hotspot and an FDTD Maxwell solver.

The \textbf{Hotspot and Hotspot3D} initial implementations used in this work are from the Rodinia benchmark suite~\cite{rodinia1_5306797}. Hotspot is a modeling tool for thermal analysis of Very Large-Scale Integration (VLSI) systems, enabling early analysis of thermal properties~\cite{hotspot_1650228}.

The \textbf{FDTD Maxwell solver} is developed as part of this work\footnote{FDTD Maxwell solver: \url{https://github.com/Jonah-E/FDTD}}. It solves the 3D Maxwell equations for a specified number of time steps for a 3D electromagnetics cavity problem. This application is slightly different compared to the skeleton application in that it launches two kernels, one for the H-Field calculations and one for the E-Field.

All time and memory measurements in the skeleton application were performed ten times for each configuration and the mean and standard deviation are shown in the figures. For the speed-up test, the time is measured for the software parts that differ between the graph and the baseline versions, e.g., the graph creation and graph/kernel launch and execution. The speed-up is then calculated as the ratio between the mean value for the baseline version and the graph version. The error for the speed-up for an iteration batch size of $S$ is calculated as the square root of the summed squares of standard deviations divided by the corresponding mean value.

Our test bed system consists of an Nvidia A100 GPU with 40 GB global memory with an AMD EPIC 7302P  as the host. It runs CentOS Linux 8 (Kernel: 4.18.0) with CUDA 12.3 with GPU Driver version 545.23.08, Nsight Systems 2023.3.3.42, nvidia-smi 545.23.08, nvcc v12.3.52, and gcc v8.5.0. We use NVIDIA Nsight Systems and System Management Interface (nvidia-smi) for collecting performance and related metrics for the NVIDIA system.

While most of the experiments are carried out on an Nvidia A100, we also test the iterative applications on a Grace-Hopper system consisting of a Grace CPU (72-core ARM Neoverse V2 CPU), and an Nvidia H100
GPU with 96 GB HBM3 memory. This system runs RHEL Linux 9.3 (Kernel: 5.14.0) with CUDA 12.2.

\begin{figure}[ht]
    \begin{minipage}{\linewidth}
        \centering
        \includegraphics[width=0.8\linewidth]{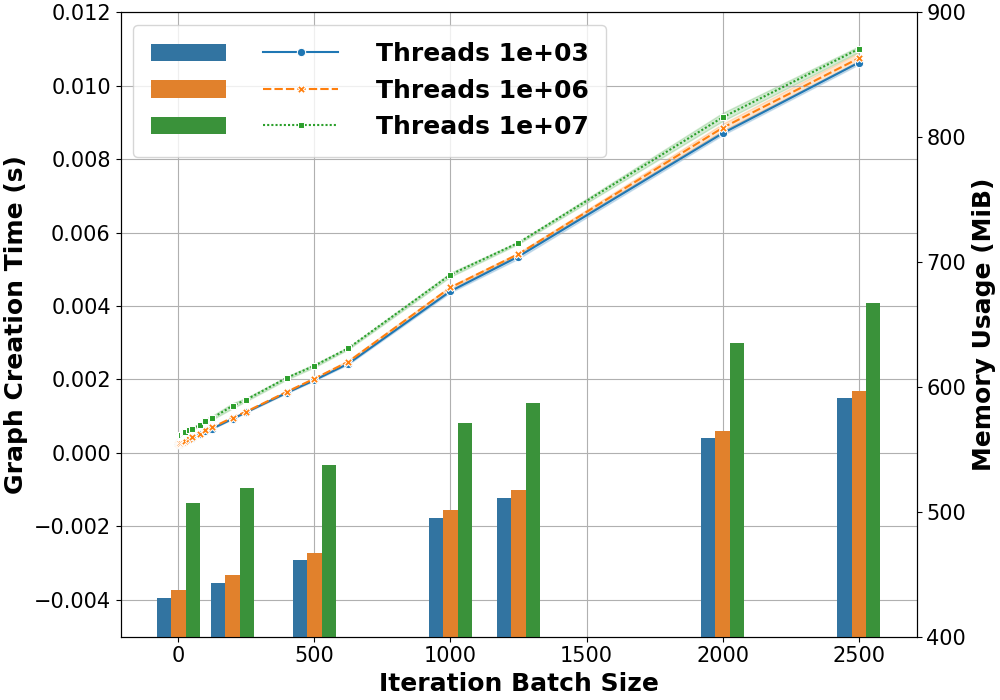}
        \caption{Graph creation overhead and memory usage in the skeleton application as the size of the iteration batch increases. The creation overhead is the line plot with the y-axis on the left side and the memory usage is the bar plot with the y-axis on the right.}
        \label{fig:mb-increase-nodes}
        \vspace{0.2em}
    \end{minipage}
    \begin{minipage}{\linewidth}
        \centering
        \includegraphics[width=0.9\linewidth]{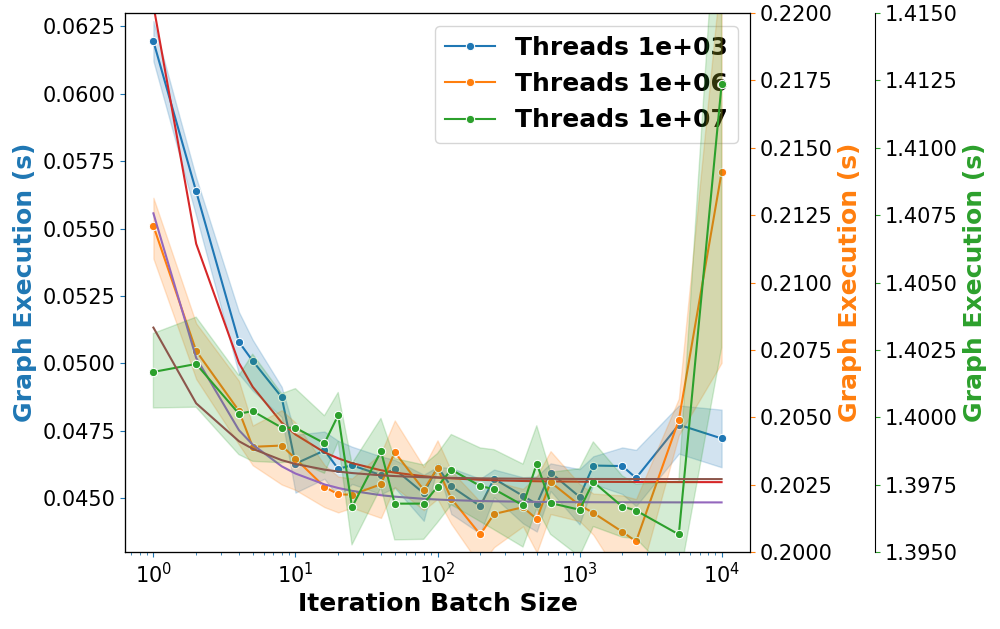}
        \caption{Launch and execution time for the graph version of the skeleton application as the size of the iteration batch increases. Note the logarithmic scale of the x-axis, and that each workload has a separate y-axis.}
        \label{fig:mb-increase-it}
    \end{minipage}
\end{figure}

\begin{figure}[ht]
    \begin{minipage}{\linewidth}
        \centering
        \includegraphics[width=0.8\linewidth]{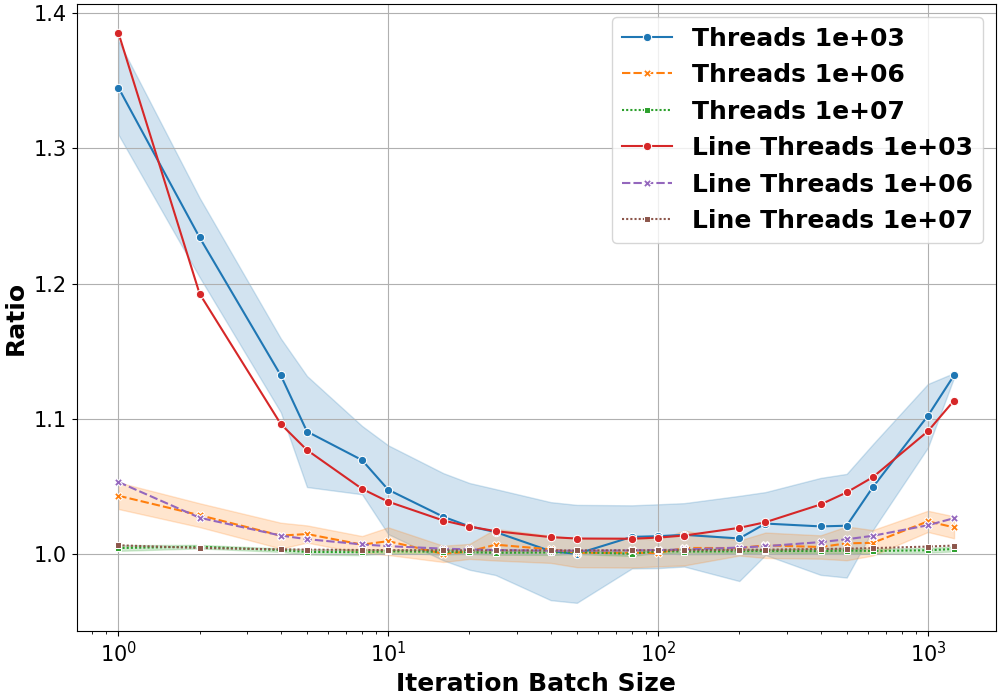}
        \caption{Skeleton application performance for different iteration batch sizes and 10,000 kernel iterations. The ratio between the graph creation time plus the total execution time for each graph size and the graph size with the lowest mean execution time. Note the logarithmic scale of the x-axis.}
        \label{fig:mb-batch-test}
        \vspace{-0.5em}
    \end{minipage}
\end{figure}

\section{Results}\label{sec:evaluation}
In this section, we evaluate the overhead and execution of unrolling iteratively launched kernels into CUDA graphs using the skeleton application. We then find the optimal size for the iteration batch and evaluate the speed-up using the skeleton application. Finally, using the results from these evaluations, we will show the speed-up for Hotspot and Hotspot3D from the Rodinia benchmark suite and an FDTD application.

\begin{figure*}[ht]
  \centering
  \includegraphics[width=0.8\linewidth]{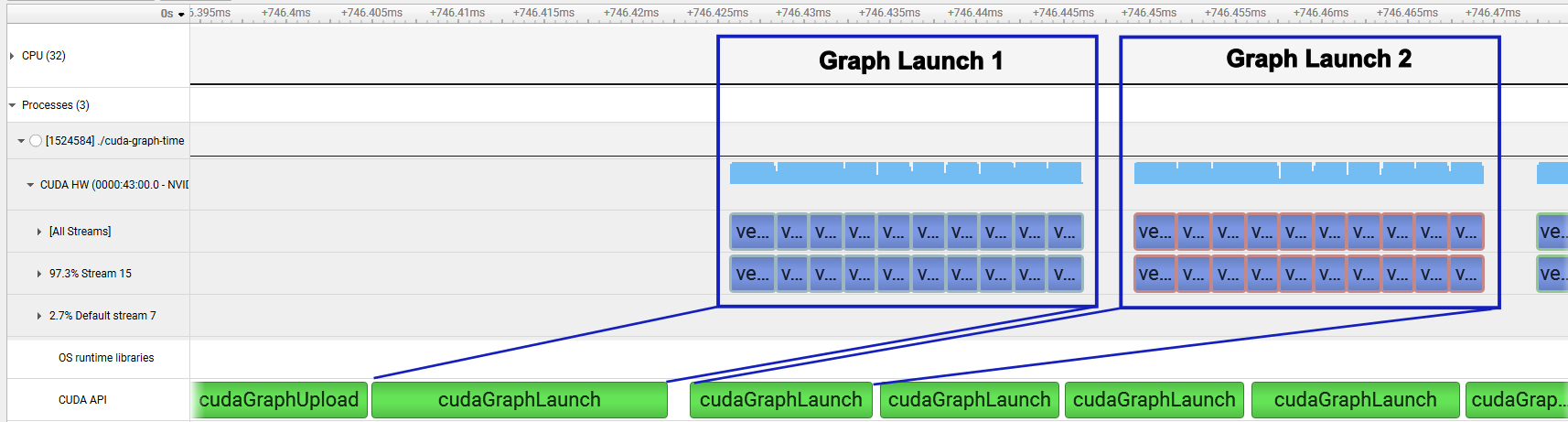}
  \caption{Execution trace from Nsight Systems. The first two graphs executed on the GPU are enclosed in boxes.}
  \label{fig:cuda-exec}
\end{figure*}

\subsection{Skeleton Application: Overhead and Execution}\label{sec:mb-overhead}
By increasing the iteration batch size, we can measure the graph creation overhead as a function of the number of graph nodes in the graph. This also includes uploading the graph to the GPU. In Fig.~\ref{fig:mb-increase-nodes} we can see that the graph creation time follows close to a linear increase with respect to the size of the iteration batch up to a size of 2,500. The coefficients $k_c$ and $b_c$ from Eq.~\ref{eq:creation_overhead} are listed in Table~\ref{tab:fit_creation}, together with the Mean Absolute Error (MEA). The slope for the creation time is similar between all three workloads, which is in line with our expectations. Above 2,500, the graph creation time does not follow the same linear increase; however, the limitation in possible iteration batch sizes from Eq.~\ref{eq:graph_executions} prevents the evaluation of alternative correlations.
Interestingly, the creation time for the largest workload, (Threads = 1e+07) has a slightly higher base time than the other two, also seen in the $b_c$ value in Table~\ref{tab:fit_creation}. We observe a similar offset in the GPU memory usage for the different workloads in Fig.~\ref{fig:mb-increase-nodes}, indicating that this increase in creation time is associated with the graph upload to the GPU.

\begin{table}[ht]
    \centering
    \caption{Fit parameters and MAE}
    \begin{tabular}{c|c c c | c c c }
                &  \multicolumn{3}{c|}{Graph Creation, Eq~\ref{eq:creation_overhead}}& \multicolumn{3}{c}{Graph Execution, Eq.~\ref{eq:exec_time_fit}} \\
        Threads & $k_c$ & $b_c$ & MAE  & $a$ & $b$ & MAE \\\hline
        1e3 & 4.18e-6 & 1.59e-4 & 9.41e-5 & 1.77e-2 & 4.56e-2 & 1.3e-3\\
        1e6 & 4.23e-6 & 1.72e-4 & 9.72e-5 & 1.07e-2 & 2.02e-1 & 1.5e-3\\
        1e7 & 4.27e-6 & 4.22e-4 & 7.28e-5 & 5.62e-3 & 1.40e+0 & 2.1e-3\\
    \end{tabular}
    \label{tab:fit_creation}
    \label{tab:fit_exectuion}
\end{table}

Using Nsight Systems, we can collect an execution trace of the graph execution in the skeleton application, Fig.~\ref{fig:cuda-exec}. Here, we can see that the assumption for Eq.~\ref{eq:graph_executions}, that the time between graph executions is longer than the time between kernel executions inside the graph ($T_i < T_a$), holds true.
\observation{\textit{The execution trace from Nsight Systems aligns with the expectation that $T_i < T_a$, presented in Section~\ref{sec:r-mb}.}}

Fig.~\ref{fig:mb-increase-it} shows how the execution time $T_e$ changes as the size of the iteration batch increases, including lines fitted to Eq.~\ref{eq:exec_time_fit}, with coefficient values in Table~\ref{tab:fit_exectuion}. Each of the workloads has its y-axis with different offsets but the same scale. As expected, the $b$ values are different for the different workloads. Unexpectedly, $a$ has different values for the different workloads. This suggests that the difference between the time between graphs and the time between kernel executions within graphs is smaller for larger workloads.

Similar to the graph creation time in Fig.~\ref{fig:mb-increase-nodes}, the execution time in Fig.~\ref{fig:mb-increase-it} does not follow the model for iteration batch sizes larger than 2,500 kernels. Instead of converging towards a constant value, the execution time increases as the size of the iteration batch is 5,000 or 10,000, that is containing either 50\% or 100\% of the total kernel executions $I_k$.

\observation{\textit{Larger workloads have lower potential performance gain from CUDA Graph.}}

To evaluate the optimal graph size and minimize Eqs.~\ref{eq:T_total}, we can measure both the graph creation and execution time together. In Fig.~\ref{fig:mb-batch-test}, we can see this together with Eq.~\ref{eq:T_total} using the fitted parameters for Eq.~\ref{eq:creation_overhead} and \ref{eq:exec_time_fit}. The data and fitted lines have been divided with the lowest mean time for each workload. From this, it can be seen that the optimal size for the iteration batch is around 50-100 kernels. However, the optimal graph size is less important when the workload is larger. The reason for this is shown in Table~\ref{tab:fit_exectuion}. First, the $b$ coefficient is larger for the larger workloads, meaning that the speed-up for increasing the iteration batch size will be lower in proportion to the total execution time. Second, the $a$ coefficient is smaller, indicating lower possible speed-ups for the larger workloads.

The memory usage should also be considered when selecting the optimal iteration batch size. Inspecting  Fig.~\ref{fig:mb-increase-nodes}, the memory usage increases linearly with the number of nodes in the graph and agrees with results from previous works~\cite{linEfficientGPUComputation2021,zhengGrapePracticalEfficient2023}. This is the overall memory usage of the GPU, we therefore cannot evaluate the absolute values but only analyze the trend for increasing graph size. Based on this, we can conclude that the optimal iteration batch size is approximately 100 nodes. Using a larger graph size will mean less optimal performance and higher memory usage.

\begin{figure}[bt]
    \centering
    \includegraphics[width=0.8\linewidth]{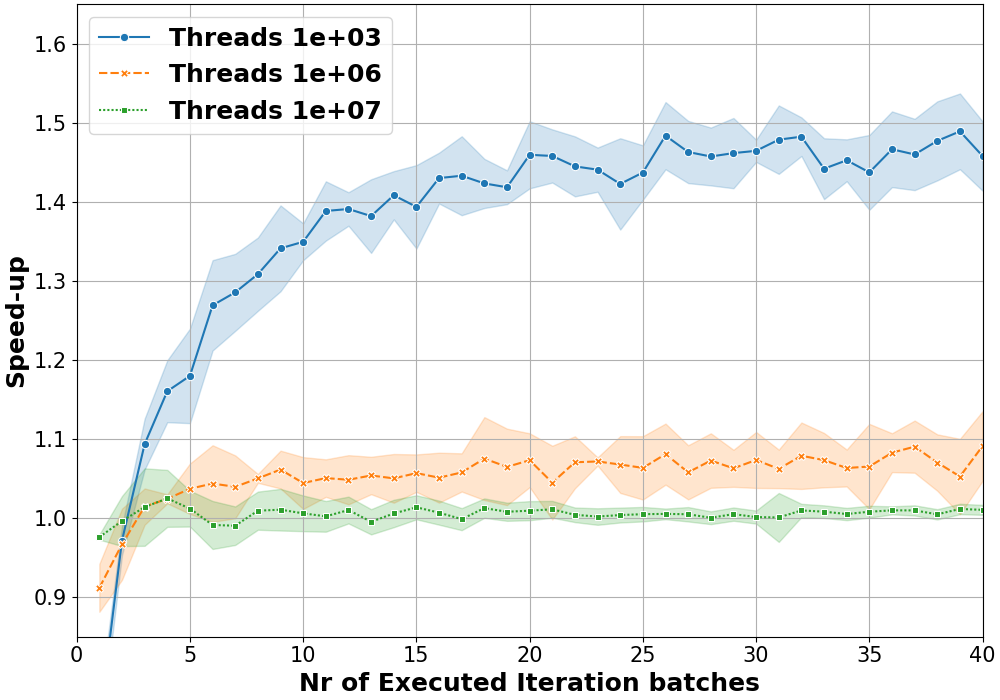}
    \caption{Skeleton application speedup of the graph version compared to the baseline version without graph.}
    \label{fig:mb-relative-speedup}
\end{figure}
\begin{figure}[bt]
    \centering
    \includegraphics[width=0.8\linewidth]{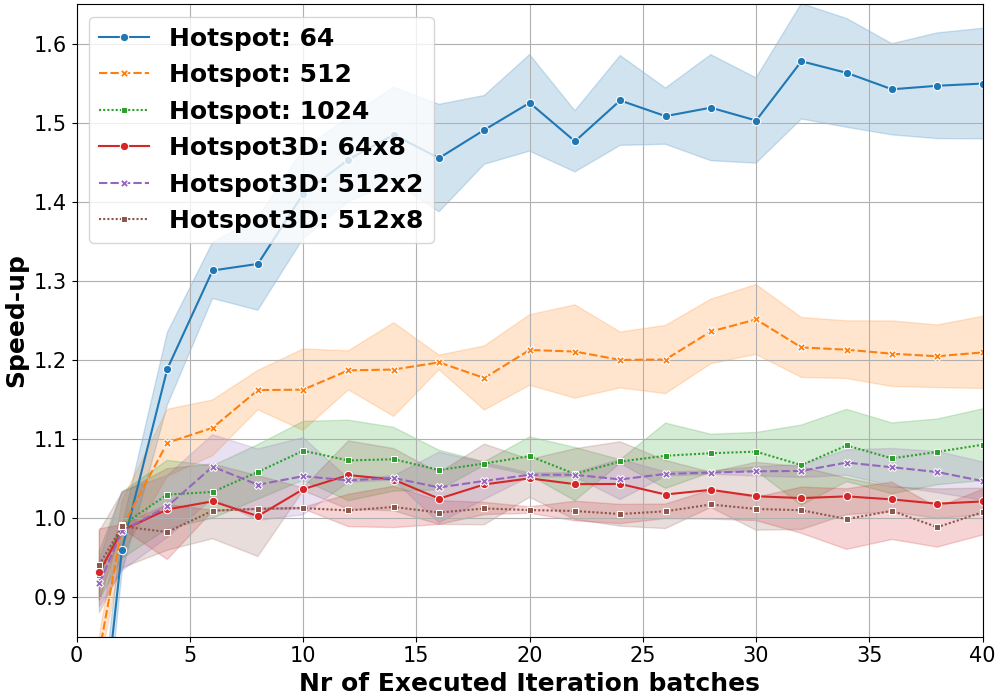}
    \caption{Speed-up for Hotspot and Hotspot3D when using \texttt{CUDA~Graph} on the A100 System. The first number for all the legend entries is the grid size, for the calculation grid. The same size is used for both the number of columns and rows in the grid. Hotspot3D also has multiple layers, the number of layers are written after the x in the legend.}
    \label{fig:hotspot-speedup}
\end{figure}

\subsection{Performance}
\noindent \textbf{Skeleton Application.} With the optimal iteration batch size from Fig.~\ref{fig:mb-batch-test} and using Eq.~\ref{eq:graph_executions} we can calculate the number of iteration batches to execute for the required kernel executions as $I = \frac{I_k}{100}$. Using this, we examine the performance gain for different numbers of required kernel launches. Here we measure the graph creation time and the total execution time of all the launched graphs/kernels, we exclude common setup steps, such as allocating memory in the GPU, from the measurement. Fig.~\ref{fig:mb-relative-speedup} shows the speed-up of the graph version of the skeleton application relative to the baseline version (without CUDA graph). We can see that for a low number of executed iteration batches, there is no speed-up. However, there is a crossover point at two to three iteration batches where the graph version becomes faster compared to the baseline version. As expected from the $a$ coefficient in Table~\ref{tab:fit_exectuion}, the smallest workloads have the highest possible performance gain with speed-up over 1.5 times, while the higher the workload the lower the speed-up.

\noindent\textbf{Hotspot and Hotspot3D.} Using our iteration batch unrolling strategy on the Hotspot and Hotspot3D benchmarks from Rodinia, the speedup, shown in Fig.~\ref{fig:hotspot-speedup}, has the same profile as the skeleton application. The larger the problem size is, the smaller the speedup is compared to the total execution time. However, after $\sim$3 iteration batches all workloads are above one, indicating that there at least is not a performance penalty.

\begin{figure}[bt]
    \centering
    \includegraphics[width=0.8\linewidth]{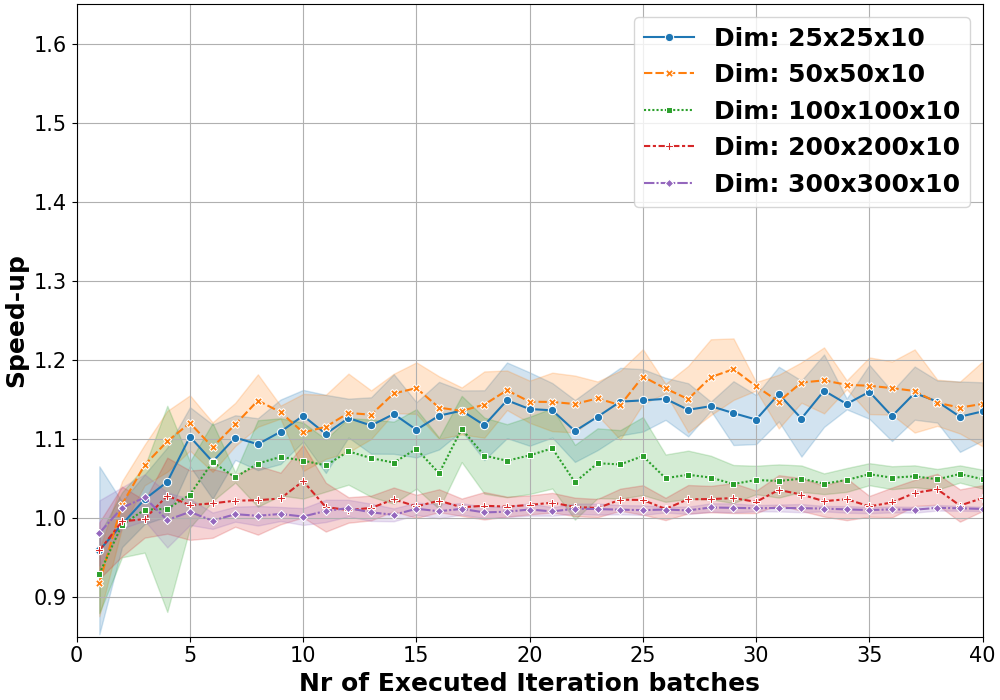}
    \caption{Speed-up for the FDTD Maxwell solver on the A100 system.}
    \label{fig:fdtd-speedup}
\end{figure}

\noindent \textbf{FDTD Maxwell Solver.} Similar to both the skeleton and Hotspot applications, a speed-up in the FDTD Maxwell solver, Fig.~\ref{fig:fdtd-speedup}, is visible. Exhibiting a speed-up of up to 1.2$\times$ for the smaller workloads tested and no slowdown for larger workloads.

\begin{figure}[bt]
    \centering
    \includegraphics[width=0.8\linewidth]{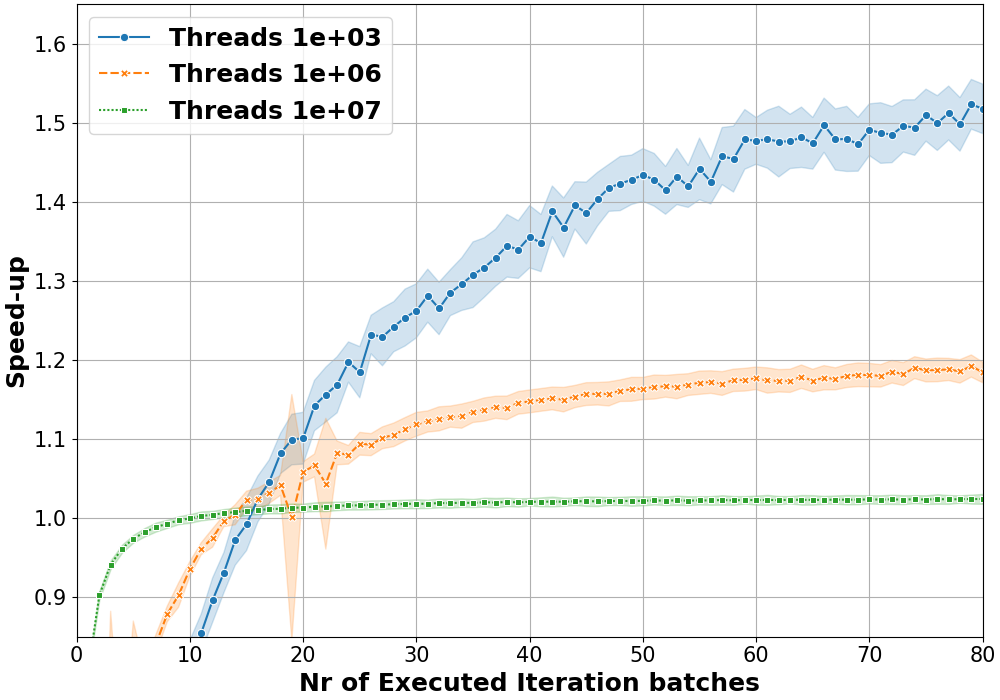}
    \caption{Skeleton application speedup of the graph version compared to the baseline version without graph on an Nvidia Grace-Hopper system.}
    \label{fig:sleipner-mb-speedup}
\end{figure}

\noindent \textbf{Execution on an Nvidia Grace-Hopper System.} Executing the skeleton application on the NVIDIA Grace Hopper system, Fig.~\ref{fig:sleipner-mb-speedup}, we observe the same general trend for the different workloads as on the A100 system, Fig.~\ref{fig:mb-relative-speedup}. However, the crossover point has moved forward, requiring a larger number of iteration batches for the \texttt{CUDA~Graph} version to be faster.

\observation{\textit{The tests using the skeleton application, Hotspot, Hotspot3D and the FDTD Maxwell solver show that the iteration batch unrolling strategy can bring more than 1.4$\times$ speed up for small workloads and have no performance penalty for larger workloads.}}


\section{Related Work}\label{sec:related_work}

\noindent \textbf{Heterogeneous Programming:} Prior works explored \texttt{CUDA~Graph} in heterogeneous programming through OpenMP and OpenACC. \textcite{yuOpenMPCUDAGraphs2020} proposed an approach to transform OpenMP task dependency graphs into \texttt{CUDA~Graph}, enabling the user to obtain the performance benefits from \texttt{CUDA~Graph} while keeping the programmability of OpenMP. 
\textcite{toledoEnhancingCodingProductivity2022} proposes an integration of \texttt{CUDA~Graph} into OpenACC, they present the benefits of this by using the Particle Swarm Optimization and integrating OpenACC-generated kernels into a \texttt{CUDA~Graph}. Our work focuses on one code pattern where one kernel is launched multiple times; the kernel unrolling into \texttt{CUDA~Graph} presented in this work could be done through directives such as those proposed by \textcite{yuOpenMPCUDAGraphs2020} and \textcite{toledoEnhancingCodingProductivity2022}.

\noindent \textbf{Performance Improvement:} Other works improve the performance of a given task graph. \textcite{qiaoBestBothWorlds2020} introduced \texttt{CUDA~Graph} to an image processing DSL and proposed a kernel pipeline approach to increase parallelism to optimize a graph based on resource usage.
CudaFlow~\cite{linEfficientGPUComputation2021} is an abstraction layer to simplify the usage of \texttt{CUDA~Graph}; with the stream capture abstraction, they also introduce a scheduling algorithm to improve the performance. 
Grape~\cite{zhengGrapePracticalEfficient2023} is a graph compiler integrated into PyTorch to improve the efficiency of graph-based execution for dynamic DNNs on GPU. Grape also introduces control flow logic for the graphs in the framework but makes no use of conditional nodes.
AutoGraph~\cite{zhaoAutoGraphOptimizingDNN2023}, is a framework for optimizing the computation graph of Deep Neural Networks (DNN) and utilizes \texttt{CUDA~Graph} to reduce the overall launch cost. Opara~\cite{chenOparaExploitingOperator2023} is a framework designed to increase the parallel execution of operators in DNN workloads. It creates a \texttt{CUDA~Graph} based on profiling information of resource usage. Regarding using \texttt{CUDA~Graph} in applications, \textcite{tangSpeechNetWeaklySupervised2022} utilize \texttt{CUDA~Graph} for speech recognition running in a commercial setting. In contrast to these works focusing on the optimization of the computation by increasing the parallelism in the execution, our work focuses on optimization without any parallelism where the full workflow is too large to efficiently express as a graph. \textcite{linEfficientGPUComputation2021} did show that the execution of a linear chain of nodes can be optimized for some performance gain over the manually defined \texttt{CUDA~Graph} for graphs with over $10^4$ nodes. However, they did not compare this to the performance without \texttt{CUDA~Graph}.

\section{Discussion and Conclusion} \label{sec:conclusions}
In this work, we presented a strategy to increase the performance of iteratively launched CUDA kernels by grouping the kernel launches into iteration batches and then unrolling the iteration batch into a \texttt{CUDA~Graph}. The basis of this strategy is the expectation that the CUDA runtime will be faster at switching to the next node inside a graph than switching independently launched workloads. We developed a skeleton application and used it to quantify the overhead and possible performance gain with the strategy. Using the same skeleton application, we showed that using an optimal size for the iteration batch, over 1.4$\times$ speed-ups is achieved. Furthermore, we showed similar speedups in other applications and on different NVIDIA GPU architectures. 

We showed that the graph creation time increases according to Eq.~\ref{eq:creation_overhead} up to an iteration batch size of 2,500. This means that for the tested $I_k$, 25\% of the number of nodes is contained in the graph. According to the model, Eq.~\ref{eq:exec_time_2}, increasing the iteration batch further would only constitute a minor decrease in execution time. The measurements of the actual execution time show that it is even worse, as the execution time increases instead of converges, as the iteration batch size increases over 2,500. However, we also see in Fig.~\ref{fig:mb-increase-it} that this is less pronounced for the smallest workload, indicating that this can be connected to the resource usage of the graph. 

With the skeleton application, we can use the same-sized graph to increase the performance for all the workloads, even the large problem sizes where the individual kernel execution is longer. This is primarily due to the graph creation time is only marginally affected by the problem size, while there is only a small difference in the performance profile between the different workloads. Increasing the workload therefore only marginally affects the absolute performance gain from using \texttt{CUDA~Graph}. However, as we have seen, the larger workloads also mean that the possible performance gain from using \texttt{CUDA~Graph} is a smaller part of the total execution.

Converting the Hotspot and Hotspot3D applications from Rodinia as well as the FDTD Maxwell solver shows that this strategy of unrolling iteration batches into \texttt{CUDA~Graph} can also be used on real applications. Where a speed-up of over 1.4 is achieved for the smaller workloads. The performance in these applications further highlights the effect the workload size has on the performance. If an application often executes an iterative loop with large workloads, it will not provide a large performance benefit to convert the application to use \texttt{CUDA~Graph}. However, if the application has small workloads, kernels with short execution times and small thread counts, it can provide a significant benefit to convert the execution to \texttt{CUDA~Graph} using our strategy to unroll the iteration batch. Even if the application occasionally executes a large workload, the strategy will not incur any penalty.


As shown in this work, \texttt{CUDA~Graph} can be used to increase the performance of iterative applications. Three weaknesses of the methodology presented are the total iterations of the kernels are limited to be a multiple of the iteration batch size, the use of manual graph creation which can be challenging to implement and maintain and it is not portable to systems without CUDA. The first weakness can be addressed using loop peeling. To address the second and third weaknesses, future work will include the iteration batch unrolling strategy into a DSL or heterogeneous programming framework, where the graph creation can be handled automatically. This could also include automatic characterization and selecting the appropriate iteration batch size for optimal performance.


\printbibliography

\end{document}